\documentstyle[preprint,aps,prb,epsfig]{revtex}
\begin{document}

\title{Itinerancy-localization duality of quasiparticles revealed by strong correlation}

\author{Byung Gyu Chae}

\address{Basic Research Laboratory, ETRI, Daejeon 305-350, South Korea}

\maketitle{}

\begin{abstract}
  The strong interaction between electrons reveals the duality of the itinerancy and the localization of quasiparticles.
The physical phenomena corresponding to each component of the duality could be realized
and coexist within the category of the uncertainty principle of the carrier dynamics,
which can be a strong reason for the complexity appearing in the strongly correlated system.
A possible mechanism for the high-temperature superconductivity is proposed on the basis of the interplay between
the renormalized expectation quantities of both parts.

\end{abstract}

  The dynamics of interacting electrons in solids is expected to be very different from
that of free electrons, and
nevertheless, the quasiparticle concept from the Fermi liquid theory has been used as a successful tool
in describing the elementary excitation of the interacting system.\cite{1,2}
However, in the strong correlation regime,
there appears to be anomalous properties away from the Fermi liquid behaviors,\cite{3,4}
so that it remains to be a question whether this description can be valid for
explaining the interaction to some extent.
Up to now,
many approaches\cite{5,6,7} within this scheme have been carried out
in order to analyze the strong correlation effects
on the electrodynamics of the carriers.

  In general, the interaction between electrons impedes their motion.
Quasiparticles in the interacting system evolve through a non-stationary eigenstate,
which indicates a weakly localized state with a damping rate.
Correlated metal sometimes exhibits localized property together with itinerant behavior,
and at the ultimate interaction,
it becomes an insulator with fully localized carriers, known as the Mott insulator.
Spectral representation of the propagator for interacting systems shows a duality
of the quasiparticle state and the incoherent background fully localized in space.\cite{8,9}
In the strongly correlated regime,
the incoherent background, as well as the weak localization of the quasiparticle state,
can be meaningful and can play a role in physical phenomenon.
We note that a strong correlation reveals a duality of the localization
and itinerancy of the charge carrier dynamics,
and it is believed that the correlated system can be explained by considering the duality component simultaneously.

  Let us investigate the carrier dynamics by using the propagator,
which is represented as the Green function.
The single-particle Green function has a general form,
$G(xt,x't') = -i \langle \Psi_{\rm{o}}|T[\hat{\psi}(xt)\hat{\psi}^{\dagger}(x't')]|\Psi_{\rm{o}} \rangle$,
where the $\hat{\psi}(xt)$ and $\hat{\psi}^{\dagger}(x't')$
are the field operators that create (annihilate) the electron.
By taking the Fourier transform, the Green function is expressed in energy-momentum representation.

\begin{equation} \label{eq:eps} G(k, \omega) = \sum_{m} \frac{|\langle \Psi_{\rm{o}}|\hat{\psi}|\Psi_{m} \rangle |^{2} }
{\omega - \epsilon_{k} \pm i \eta} \end{equation}

  The spectral density function is obtained from the identity,
$A(k,\omega) = -(1/\pi)\rm{Im} G(k,\omega)$,
resulting in $\sum_{m} |\langle \Psi_{\rm{o}}|\hat{\psi}|\Psi_{m} \rangle |^{2} \delta (\omega - \epsilon_{k})$.
In a non-interacting system,
the Green function has one pole with free particle energy $\epsilon_{k}$,
where the spectral density is delta-function.
But, the interacting system has spectral function that spreads out from the superposition of many poles.
The representation of the Green function in time-domain is favorable for investigating the itinerancy of the quasiparticle,
$G(k, t) = \int G(k, \omega) e^{-i\omega t} d \omega$.
By the contour integral,
the function can be split into a coherent quasiparticle state and an incoherent background with no pole.\cite{2}

\begin{equation} \label{eq:eps} G(k, t) = -iZ_{k} e^{-i\varepsilon_{k}t-\Gamma_{k} t} + G_{inch} \end{equation}

  The first part of the equation shows that the state propagates like a wave packet
with an energy $\varepsilon_{k}$ and damping rate $\Gamma_{k}$.
The amplitude of the wave packet is the residue of the function,
known as the renormalization constant $Z_{k}$.
Figure 1(a) depicts the dynamics of the quasiparticle in the lattice model that has the strong interaction.
In general, the overlap of the atomic orbitals in lattices removes
the degeneracy of each orbital to form a band,
which makes it possible for the electron to freely migrate within the band.
As shown in Fig. 1(a), in the case of a weak overlap of the orbital,
the electron is no longer a free particle and rather, feels the potential barrier to move around,
where the carrier state is the wave packet that is slightly localized in space.

  The Green function in the interacting system can be expressed in terms of the self-energy
$\Sigma(k, \omega)$, as like $G(k, \omega) = 1 / (\omega - \varepsilon_{k} - \Sigma(k, \omega))$.\cite{10}
From this notation, the renormalization constant and the elementary excitation energy are expressed
as $Z_{k} = (1 - \partial \rm{Re}\Sigma)^{-1}$ and $\varepsilon_{k} = Z_{k} (\epsilon_{k} + \rm{Re}\Sigma)$,
respectively.
In Fig. 1(b),
the spectral function shows a Lorentzian form with a height $Z_{k}$ and width $\Gamma_{k}$,
where the latter is proportional to
$[(\pi k_{B} T)^2 + (\varepsilon_{k} - \epsilon_{F})^2]$.
The Fermi liquid description is valid when it is close to the Fermi surface.
On the other hand,
when the renormalization constant approaches zero,
the quasiparticle is considered to be fully localized in the lattice site.
In Eq. (2),
only the incoherent part of the Green function remains at the zero value of the renormalization constant.
It is ensured that $G_{inch}$ is associated with the localization of the quasiparticle
in the lattice site.

  The strong many-body interaction increases the self-energy
and causes the elementary excitation energy to not be a negligible value.
The quasiparticle may not be a good, effective single-particle state,
and in addition, the incoherent part of the Green function has a value
that cannot be ignored.
In this situation,
the interacting particle can behave in two ways;
one is a mobile quasiparticle state and the other is an incoherent background localized in the lattice site.
In the limit t $\to$ 0, the Green function presents a momentum distribution in Fig. 1(c),
where $Z$ becomes the discontinuity at the Fermi surface.\cite{11}
We consider that the incoherent part of the Green function is closely related to the particle-hole excitation,
from the curve of the momentum distribution.
This is depicted in a dotted line in Fig. 1(b),
which can be a strong clue of the collective excitation in the correlated system.
This is a different concept from the collective modes derived by
the instability of the nesting vector of the mobile carriers
or in the Luttinger liquid in one dimension.

  According to the sum rules,
the spectral density function is represented as the quasiparticle part,
$\int A_{ch}(k, \omega) d \omega = Z$
and the incoherent background, $\int A_{inch}(k, \omega) d \omega = 1-Z$.
The value $Z$ becomes a measure to evaluate the itinerancy of the quasiparticle.
We simply extract that the coherent part of the Green function is weighted by the itinerancy factor $Z$,
and the incoherent part by the factor $1 - Z$,
because $A(k, \omega)$ is only the probability of adding and removing particle to the many-body system.
Namely, the Green function has the form, $G = Z G'_{ch} + (1 - Z) G'_{inch}$.

  The interacting system is described by the basic Hamiltonian
with a kinetic energy and Coulomb energy.
We infer that the Hamiltonian can be clearly divided into coherent and incoherent parts
because both parts have independent eigensates in time.
Considering that the physical observable is represented as the expectation quantity,
it is convenient to assume that
the operators of both parts are themselves weighted by factors $\sqrt{Z}$ and $\sqrt{1 - Z}$, respectively,
which imply the weight factor, as like $c^{\dagger}_{k} = \sqrt{Z} c'^{\dagger}_{k}$.
In the coherent part described by operators,
$c^{\dagger}$ and $c_{k}$,
the state propagates with a quasiparticle eigenvalue $\varepsilon_{k}$ and weighting factor $Z$.
Effective Hamiltonians $H_{ch}$ of coherent part is expressed as follows,

\begin{equation} \label{eq:eps} H_{ch} = \sum_{k} \varepsilon_{k} c^{\dagger}_{k} c_{k} +
\sum_{k,k',q} V_{q} c^{\dagger}_{k+q} c^{\dagger}_{k'-q} c_{k'} c_{k}  \end{equation}

where $\varepsilon_{k}$ and $V_{q}$ are the dressed kinetic energy
and interaction between quasiparticles.
In a similar way,
the operators of incoherent part
can describe the collective excitation with a weighting factor $1-Z$.
Effective Hamiltonian $H_{inch}$ corresponds to the Hubbard model except for the electron localization,\cite{12,13}
where the electron can not move to the nearest site due to the on-site interaction $U$
larger than the hopping energy $t$.
The hopping energy plays a role only in the spread of the excitation bands,
so that this may well describe the collective excitation of incoherent bands, in Fig. 1(b).

  From the above propagator formalism,
we find that the many-body correlation reveals a duality in the itinerancy
and localization of the quasiparticle in the strongly correlated system.
This is based on the wave-particle duality in the basic concept of quantum mechanics.
However, there is a crucial point here, namely the duality arises from not an external measuring operation
but a strong correlation in the system.
It is very interesting to know whether the physical phenomenon corresponding to
each component of the duality could be simultaneously realized and coexist in the system,
and further, interplays with each other to make a new phenomenon.

  Considering the splitting of the coherent and incoherent states,
the instability exists in knowing which component is dominantly revealed in the system,
especially when two components have comparable values.
This closely resembles the observing process in quantum phenomenon,
that is, the physical quantity in possible states can be uncovered
only after an external observing process.
Here, we can guess the spatial extent of the quasiparticle state
from the uncertainty principle,
$ \Delta x \Delta p \geq \frac{\hbar}{2} $.
The coherent part of the propagator is obtained from the summation of the possible excited states
of an adding particle and thus,
the particle-like peak spreads out in the momentum space.
Namely,
the plane wave transforms into a wave packet slightly localized in space,
and then evolves over a lifetime.
Assuming that the lifetime of the quasiparticle is about several decades meV,
the ability to confirm the location of the quasiparticle state extends over several decades \AA.
On the other hand,
as mentioned previously, the incoherent background can make the collective excitation.
Therefore, electronic inhomogenity will appear in the strongly correlated system.

  Our approach is applicable to the understanding of the high-temperature superconductor
known as the representative of the strongly correlated system.
The parent material of this system is the Mott insulator with antiferromagnetic ordering,
and by deviating from the half-filling, collective excitations appear.\cite{4}
Elementary excitation is generally analyzed by using the Hubbard model in the mean-field basis.\cite{14}
However, this analysis does not consider the renormalization factor.
Our study describes the superconductor and density wave parameters, $\Delta_{SC}$ and $\Delta_{DW}$
as the renormalized quantities by the itinerancy factor $Z$.

  Doping of the holes (electrons) to the correlated system leads to the complexity
presumably including various collective modes and in this regime, electronic inhomogenity appears,
which makes it difficult to analyze apparently physical properties.\cite{3,4}
According to our scheme, this can be interpreted as natural results by
the itinerancy-localization duality of the quasiparticles.
Here, the pseudogap states is a kind of collective excitations of the incoherent background.
This is in agreement with the midinfrared excitation of the optical spectroscopy,\cite{4,15}
leading to closure of the charge gap before close to the metal state.

  Another important property is the possibility of the pairing of the quasiparticles.
Charge carriers move in collective excitations of the incoherent background
that can be considered as bosonic glue.
Even under weak attraction, fermion particles at the Fermi surface are unstable to form the bound state.\cite{16}
We consider the system of electrons (holes) and bosons with mutual interaction.

\begin{equation} \label{eq:eps} H_{el-b} = \sum_{k,q} W_{q} (b^{\dagger}_{-q} + b_{q})
c^{\dagger}_{k+q} c_{k}  \end{equation}

where $W_{q}$ is the interaction matrix element and $b^{\dagger}_{-q}$ the boson operator.
The effective interaction between two electrons is obtained by canonical transformation,
which arises from the exchange of the boson such as collective modes in analogy with the phonon exchange.

\begin{equation} \label{eq:eps} V_{eff} = \sum_{k,k',q} \frac{|W_{q}|^2 \omega_{q}^{2}}{\omega^{2} - \omega_{q}^{2}}
c^{\dagger}_{k+q} c^{\dagger}_{k'-q} c_{k'} c_{k}  \end{equation}

For $\omega < \omega_{q}$, the effective interaction is attractive.
The pairing of the carrier is possible at low frequencies, leading to the superconductor.
According to the BCS formalism, the gap equation is expressed as below,

\begin{equation} \label{eq:eps} \Delta_{SC} = 2\hbar \omega_{D} \exp(-\frac{1}{NV})  \end{equation}

  where $N$ is the density of states at the Fermi surface
and $\omega_{D}$ is the cutoff frequency of the collective excitation of incoherent background.
The spin wave can be formed as a possible candidate of the collective modes,
because in Eq. (5), the strong on-site interaction with compared to the hopping energy leads to
the Heisenberg hamiltonian, so that the finite-range antiferromagnetic ordering exists.
This is very analogous to the spin-bag concept\cite{17},
but an important difference is that the expectation values of two phenomena,
the quasiparticle pairing and collective mode are linked by the itinerancy factor.
Since the gap parameter, $\Delta_{SC}$ is represented as an anomalous Green function,
$e^{-2\mu t} \langle c_{k\uparrow} c_{-k\downarrow} \rangle $, where $\mu$ is the Fermi energy,
it implies the factor $Z$ due to the symmetry of the many-body system.
The energy, $\hbar \omega_{D}$ comes from the expectation quantity of the spin wave.
Therefore, the critical temperature can be written as, $T_{c} \simeq \ Z(1-Z)$.
Figure 2 shows the change of the critical temperature with respect to the itinerancy factor $Z$.
Assuming the itinerancy factor is proportional to the doping carrier,
$T_{c}$ variation is similar to the experimental results.\cite{4}
This relation can be valid for the both states of electron and hole doping.

  In conclusion, the possibility to interpret the complexity of the strongly correlated system is proposed
within the itinerancy-localization duality of the quasiparticle.
The strong correlation makes the system reveal the duality of the quasiparticle
and new physical phenomena through the interplay between them.
This scheme is applicable to the analysis of other physical phenomena induced by the strong interaction between carriers.

\begin{figure}
\vspace{0.0cm}
\centerline{\epsfysize=18cm\epsfxsize=12cm\epsfbox{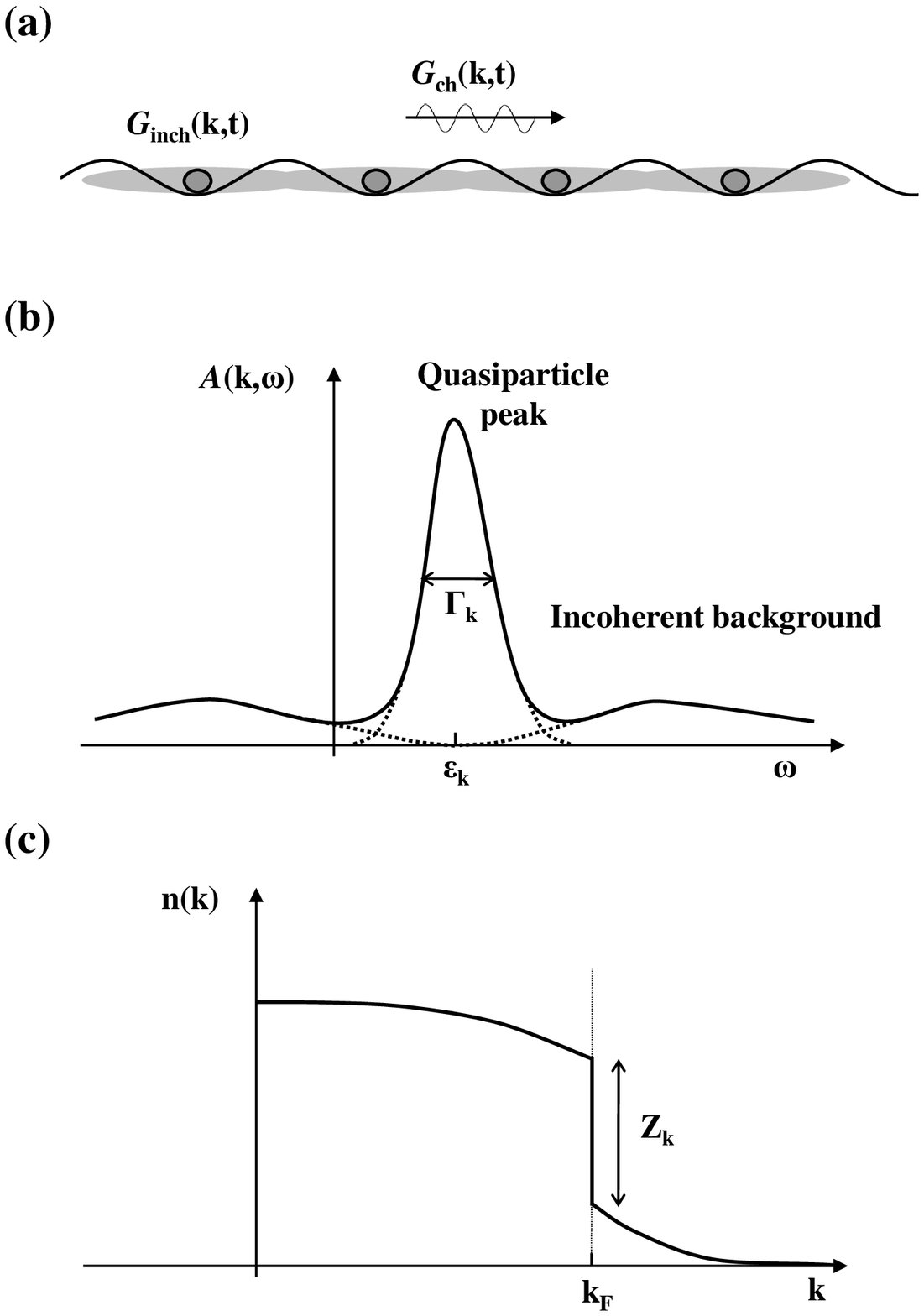}}
\vspace{-2.0cm}
\caption{(a) A schematic diagram to visualize the
duality of the itineracy and the localization of quasiparticles. The
Green function with a many-body correlation can be split into a mobile
quasiparticle state, $G_{ch}$ and the incoherent background
localized in the lattice site, $G_{inch}$. In the Fermi liquid state,
(b) the spectral density clearly shows a quasiparticle peak and incoherent background
and (c) the momentum distribution has discontinuity at the Fermi surface.}
\label{f1}
\end{figure}

\begin{figure}
\vspace{0.0cm}
\centerline{\epsfysize=15cm\epsfxsize=10cm\epsfbox{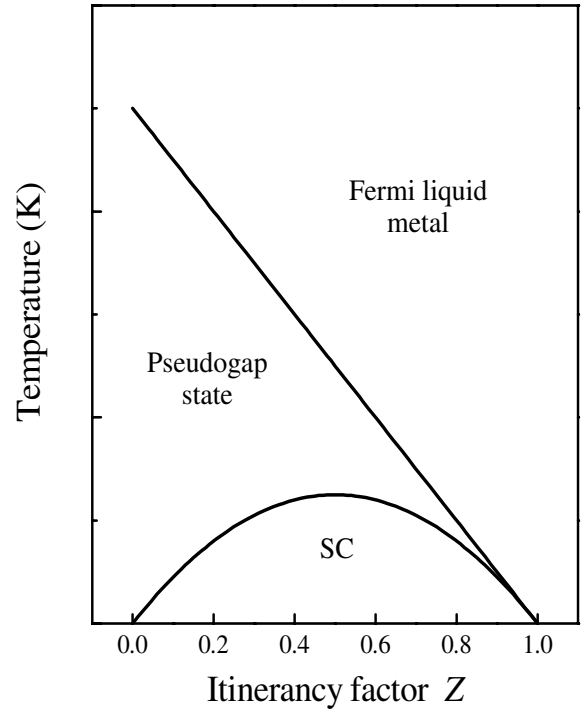}}
\vspace{-1.5cm}
\caption{Change in the critical temperature of the
superconductivity with respect to the itinerancy factor $Z$. The
straight line proportional to 1 - $Z$ indicates a pseudogap
temperature from the incoherent background.}
\label{f2}
\end{figure}

\end{document}